\documentclass[%
reprint,superscriptaddress,bibnotes,
amsmath,amssymb,aps,pra,
]{revtex4-1}

\usepackage{scrextend}
\newenvironment{transcript}
    {\begin{addmargin}[0.32cm]{0.32cm} 	     \vspace{0.15cm}}
    {\vspace{0.15cm} \end{addmargin}}
\usepackage{color}

\usepackage[T1]{fontenc}	
\usepackage[latin9]{inputenc}	
\usepackage{geometry}		
\usepackage{graphicx}
\usepackage[above,below]{placeins}	
\usepackage{times}

\usepackage{hyperref}  
\hypersetup{colorlinks=true,urlcolor=blue,citecolor=blue,linkcolor=blue}   
\begin{document}
\raggedbottom
\setlength{\parskip}{0pt}
\title{An uncommon case of relevance through everyday experiences}
\author{Abhilash Nair}
\affiliation{Department of Physics \& Astronomy,Michigan State University, 567 Wilson Rd, East Lansing, MI, 48824}
\author{Vashti Sawtelle}
\affiliation{Lyman Briggs College, Michigan State University,919 E Shaw Ln, East Lansing, MI, 48824}
\affiliation{Department of Physics \& Astronomy,Michigan State University, 567 Wilson Rd, East Lansing, MI, 48824}

\begin{abstract}
Physics education research has probed for the relevance of physics in students' everyday lives. Attitudinal and epistemological surveys have asked students if they think of or use physics in their daily lives. We have previously documented how it is uncommon that our life science students describe using or even seeing physics in their daily life \cite{Nair2018}. This result was unsurprising and aligns with previous scholarship of students majoring in disciplines outside of physics; we have argued that it is optimistic for scholars to expect students with disciplinary homes outside of physics to see their experiences through a lens of physics. Methodologically, we searched for a contrasting case (Sam). Sam is majoring in the life sciences and articulates moments where she uses physics to reason through everyday phenomena. We explore the ways in which courses can support students like Sam to find physics relevant to their everyday experiences.
\end{abstract}

\maketitle
\section{Introduction}
The Physics Education Research (PER) community commonly uses attitudinal and epistemological surveys \cite{Adams2006,Redish1998a,Halloun1998,elby1998epistemological} for a variety of reasons including the evaluation of courses as well as the measurement of students' beliefs around the relevance of physics. For example, the Colorado Learning Attitudes about Science Survey (CLASS) contains two clusters of items that probe relevance: \textit{Real-World Connection} and \textit{Personal Interest} \cite{Adams2006}. The majority of the items in these two clusters focus on how students think of or use ideas in physics outside of the physics classroom with many asking about students' everyday lives.

In our introductory physics for life sciences (IPLS) course we have found that students rarely report bringing physics ideas into their everyday experiences. Our case students have commonly laughed when we ask if they think of, use, or talk about physics in their daily life. This result is not surprising when placed in the context of previous scholarship on students beliefs around physics. 

Elliott \cite{atkins2017} reports similar stories of students laughing at the thought of bringing physics ideas outside of the classroom. We also have many studies on survey measures reporting that they typically measure a deterioration in students beliefs around physics as a result of instruction \cite{Adams2006,Redish1998a}. Nair, Irving, and Sawtelle \cite{Nair2018} have previously argued that it may be optimistic for physics instructors to expect or hope for life science students to see their everyday experiences through the lens of physics.

In this paper we complicate our previous assertions and present a contrasting case of a life science student (Sam, a pseudonym) who reports physics impacting her everyday experiences. The goals of this paper are to (1) articulate the critical features of Sam bringing physics into her everyday life and (2) discuss the implications of Sam's story in designing more relevant physics classroom experiences for life science students.

\section{Theoretical Framework}
Relevance is a term that is commonly used in calls for reforming physics instruction \cite{NAP13165,Wieman2005,Stuckey2013} and in reports on the impact of instruction on student beliefs about physics \cite{Brewe2013,Crouch2018,Bennett2016}. We operationalize relevance as a construct based on how surveys in PER have described and measured areas of relevance \cite{Adams2006,Redish1998a,Halloun1998,elby1998epistemological} and theoretical descriptions of relevance as a relation between the student and physics \cite{Bookstein2007,Newton1988}. We define relevance as \textit{the opening of a conduit between two settings or experiences, through which meaningful knowledge or skills can be exchanged}. 

In a classroom, relevance can be observed in moments where students bring in experiences from outside the classroom in meaningful ways. Reciprocally, it can also be seen as students bringing classroom experiences out of the classroom into the real world, future careers, or everyday life. Beliefs of relevance can be elicited through surveys or interviews and constitute only part of the complete picture. Another critical part of relevance is situated in and facilitated by the interactions between students and their environments. For the purposes of this paper, we limit our attention to the ways in which students see the relevance of physics through everyday experiences. We employ a situative perspective \cite{Greeno1996} to describe relevance as emerging through the systems of relationships, roles, and settings students experience \cite{Bronfenbrenner1979}.

\section{Methods}
This study is situated in the first semester of an IPLS course that is taught in the studio format. The course adapts available IPLS curricula and designs new materials with the aim of making physics relevant to life science students. This course also adapts discursive structures and participation frameworks from Modeling Instruction for introductory university physics \cite{Brewe2010,Brewe2008}. The classroom structures, norms, and participatory roles were designed to position students as experts, leveraging their extensive disciplinary experiences in biology and chemistry as they learn physics \cite{Sawtelle2016}. The course's design encourages students to engage in the practice of physics through argumentation and white-board meetings \cite{Brewe2008} in order to arrive at a classroom consensus.

The data presented here is part of a larger effort to operationalize the construct of relevance for study in PER \cite{Nair2018} as well as to articulate design conjectures of a classroom that attends to students' identity, affect, and beliefs \cite{Sawtelle2016}. To identify students for this larger data set, we solicited volunteers for interviews and in-class video observation. Volunteers were cross-matched with results from a course survey to identify volunteers who had a strong disciplinary identity, were fearful or anxious of physics, were skeptical of the value of the course, or were optimistic that the course could be relevant.

This paper focuses on data from one student, Sam. We present data from three interviews, each interview was video-recorded and followed a semi-structured protocol that included items to probe the relevance of physics to students' lives. The first interview took place in the first month of Fall semester. The second interview was conducted a month before the end of the Fall semester, and the third interview took place in the middle of the Spring semester, after Sam had moved onto a second semester physics course with different design goals.

\section{Sam}
Sam is a student dual majoring in nutritional sciences and kinesiology. Her mother is a physician and her father is a physical therapist. At the time of this study Sam is considering a range of careers paths including neurology, orthopedic medicine, and psychiatry. Sam is interested in fitness and is an avid weightlifter working towards her personal training certification. In her initial course survey Sam reports that her personal training lessons are about "how the body functions as a series of machines, so physics is part of the human body and not just simple machines." She has previously taken AP Physics and in the first interview describes herself as consistently receiving "straight A's" in high school. In college she is finding the classes more difficult and her grades have dropped. In the initial survey Sam reports ---on a scale of 1-to-10--- maximum anxiety (10/10) about the physics course, low level of preparation (3/10), and low sense of capability (4/10). We note this here to point out that Sam is not intrinsically interested in physics from the start, but in the following sections, we will describe moments in which Sam describes physics being relevant to her everyday experiences. Consistent with our definition of relevance, we frame these examples as intersections of physics and the different settings, roles, and relationships in Sam's life.

\subsection{Classroom activities intersecting with Sam's lived experiences}
Early in the Fall semester, students work through an activity on wound healing in which they are tasked with analyzing the motion of different cells moving in videos captured from a microscope. Students are asked to compare the speeds of the bacteria, neutrophils, and tissue healing to conclude whether antibiotics would be required \cite{Moore2014}. Sam found this activity interesting and related it to her own experience with a wound.

\begin{transcript}
\textbf{Sam}: The wound healing. Yeah that was very interesting... it was actually really cool to see the video of the wound closing... I had this severe cut on my arm, so I got to watch over time how the wound healed and how it's still healing right now... so watching that on video was really cool to me. To see how the body can reform itself and produce scar tissue when it's been completely sliced. [Int 1]
\end{transcript}
Sam connects the wound healing activity to her own experiences with a cut and watching it heal over time. Near the end of the first interview the first author asks Sam "Have you had any moments that you're proud of in class?"

\begin{transcript}
\textbf{Sam}: Yeah. The wound healing. I keep going back to that. Being able to track that certain point; I found very satisfying. I was very proud of myself to be able to find the rate at which that wound healed. And if I could do that on a real person that would be really cool. Tell this person, oh you need antibiotics because this wound is not healing fast enough because this travels faster. [Int 1]
\end{transcript}
This activity becomes a central experience for Sam that she recounts in every interview, even well after finishing the Fall semester. In the second interview, Sam states that this activity was her favorite and that it was "\textit{super eye-opening in how many different ways you can test whether or not a patient needs something}. [Int 2]" In the third interview, Sam describes "\textit{it was just really cool to see what actually happened in my arm}. [Int 2]" Sam also reports that she talked to people all the time about the wound healing activity telling them "\textit{it's just like what happens in everyday wounds}! [Int 3]" She then jokingly tells us that people tell her she's crazy for talking about the activity.

In this section, we have seen the wound healing activity impact Sam's sense of the relevance of physics; the activity aligned with her interests, gave her a new perspective on her own experience with wounds, and led to conversations about physics in her daily life. In addition to this classroom activity, Sam reports seeing connections to physics in her everyday experiences in the gym and discussions with her friends. In the next section, we will describe Sam making relevant connections with support from her peer network.
\newpage
\subsection{Social activities \& peers support Sam to see the world through the lens of physics}
Our first sign that Sam's peers had an impact on her making connections to physics in her daily life is in Interview 1. Sam states "I like how... this class relates everything to what this world is." We then ask her to give us more words around this.

\begin{transcript}
\textbf{Sam}: You can apply physics to pretty much everything.\\
\textbf{Interviewer}: When did you start thinking that physics can be applied to everything?\\ 
\textbf{Sam}: I have some friends who are engineers and so they're pretty much into physics I guess. When I hear them talk about, how this works in this way, that makes me think. I have a friend who's an engineer right now or who's schooling in engineering right now and he just got into weightlifting. I'm helping him out and stuff. And he's like I find it really interesting how the body is a bunch of levers and pulleys. I've never thought of it that way. And in my course for personal training, there's a whole chapter about how the body is a bunch of machines. [Int 1]
\end{transcript}
Sam's engineering friend, who is enrolled in a different physics course at the same time as her has an impact on how Sam thinks about the relevance of physics to the human body. Later in the interview, we ask "Do you ever find yourself sitting and thinking about it like outside of the classroom setting?"

\begin{transcript}
\textbf{Sam}: Uh-huh (in agreement). I go to the gym with my engineering friend a lot and doing a set either of us watching each other or just watching myself, I see how that works. You could also use physics to determine how to work a muscle better. Because if you were to ---for a biceps curl for instance--- if you were to go all the way down you would get most of the muscle and bring it all the way up. Whereas going down to 90 degrees is not going to work the entire muscle...\\
\textbf{Interviewer}: I see. Is your friend also into thinking about the levers and stuff? Or is it something that you bring in?\\
\textbf{Sam}: Yeah he is. Yeah he is the one who brought the idea into my head. He's really into physics. He loves physics and so he'll find any way to relate back. [Int 1]
\end{transcript}
This engineering friend plays an important role in promoting conversations around physics in social settings. A month away from the end of the Fall semester, we talk to Sam again and ask "When you're out in the real world, do you ever find yourself thinking about the physics you learned in physics class?"

\begin{transcript}
\textbf{Sam}: Yes, actually today, I was thinking; I was on my way biking to my 8 am class... I was on my highest gear, and I was like, what's the force I'm exerting on this peddle? ...because it's not the force on the peddle, it's like the force in my muscles. And how do my muscles get energy to produce that amount of force? Well, when it's in the food that I eat, and then where does that energy come from? It's just cycles. So sometimes it pops into my head, yeah.\\
\textbf{Interviewer}: Awesome. Were there any other examples that come to mind?\\
\textbf{Sam}: With my engineer friend, he brings it up a lot. ...we were talking about how the wheels-to-wheels law, where you're supposed to be biking with the traffic. ...we had this debate, and he's like, "Well, you're more safe if both momentums are going the same direction, as opposed to a head-on collision." And then I brought up the point where you have time to react when you're going the opposite way, whereas, if they just come up and clip, you don't have time to react because you don't even see them. So he brought up a bunch of physics... we just had the large debate and brought physics into it. [Int 2]
\end{transcript}
This debate over the correct rationale for the direction of bicycles on the road is memorable for Sam as she brings it back up in the third interview. She reiterates both her and her friend's arguments and mentions that they may still jokingly bring up this debate. We then ask "do you have other moments where you are kind of in the world, observing things and physics comes to mind?"

\begin{transcript}
\textbf{Sam}: Yeah! If you see an icicle fall, depending on how high it is, it could totally kill someone. (laughs) I was like talking about that with a friend like 2 days ago.\\
\textbf{Interviewer}: Really? Take me into that conversation.\\
\textbf{Sam}: We were walking  ---because I live in an apartment now--- and we were walking by the complexes and there are these giant icicles hanging from the gutters and she was like "Wow! Those could kill you." I was like well I guess it depends on how far they fall because if they're right above you then there's not much gravity to carry it down. And she's like well I guess so but the height they are they probably could kill you because they're huge! And that just brings in the mass and acceleration because you have the acceleration due to gravity and you have the mass of the icicle. Yeah, so that's where that conversation went.\\
\textbf{Interviewer}: Why do you think you're noticing these things? What would you credit that to?\\
\textbf{Sam}: Um...I actually would credit it to my friend, he's an engineer... because he sees physics in everyday life more than I do. And so when he'll bring it up, it'll make the wheels in my head turn. And now that I understand these concepts given my professors, I can apply that to different things that I see. So I guess I credit it to everybody. Now that I have the knowledge and someone is helping me apply it to make my mind start going off somewhere else where it normally wouldn't. [Int 3]
\end{transcript}
Sam credits her seeing physics in her daily life to both the knowledge she's gained from learning physics and her engineering friend in helping her apply those concepts to the world. She then makes connections to physics unprompted to reason through what variables are important in considering the lethality of falling icicles.

\section{Discussion \& Conclusion}
Connecting physics to everyday life is a major component of attitudinal and epistemological measures in PER that serve to evaluate course designs and their impacts on student beliefs around physics. Previous scholarship and our own findings have reported that it is likely students will leave a physics course believing physics and the real world are more disconnected than when they entered the course. This commonly found result has been reported as a negative shift or deterioration of students beliefs around physics.

We presented a contrasting case of Sam, a student who reports thinking about physics in her daily life and seeing relevant physics as she experiences the world around her. We describe the supports both in her course activities as well as her peer network that have helped facilitate and promote these connections to everyday life. In curricular reforms to make physics relevant to the life sciences, there have been efforts to make physics activities relevant to student interests by framing them in contexts that students may find interesting \cite{Redish2014,Crouch2018} or related to their future careers \cite{Bennett2016}. Sam's experiences show us that this is only part of the solution. Sam was certainly interested in the wound healing activity, but an important part to her finding that activity relevant was her own lived experiences of watching her own wound heal and her desire to understand how it worked. We also see the role of Sam's peer network, especially her engineering friend, in her seeing the world differently.

Sam's experiences suggest that future reforms in making physics instruction more relevant should leverage the opportunity to allow for students to bring their rich lived experiences into the classroom. Her stories also imply that students are capable of making connections to physics if there is sufficient support and intersection between their interests, coursework, and support systems. The relevance we see in Sam's stories highlights the importance of a situative approach \cite{Greeno1996} - the moments in which Sam finds relevant connections to physics exist at the intersection of physics and the systems in her life. It is the interaction of Sam's classroom activities, her interests, and her support systems that facilitate the co-construction of the notion that physics is relevant to Sam's everyday life.

\section{Implications for Future Work}
We have previously troubled the notion of the abstract "real world" and described how students are able to use their disciplinary experiences to make relevant real world connections to physics \cite{Nair2018}. Students' beliefs around the role of physics in the "real world" and their everyday life have constituted the bulk of the commonly used items designed to probe students' perceptions of the relevance of physics \cite{Adams2006,Redish1998a,Halloun1998,elby1998epistemological}. Research into both areas of relevance through narratives of student experiences have highlighted the importance of classroom structures, norms, and culture in helping facilitate and amplify students' sense of relevance. In future work, we set out to articulate design principles that will guide the creation of more relevant physics classrooms and push for holistic classroom reform efforts beyond applying a veneer of relevance in problem statements.

\acknowledgments{We thank the members of ANSER \& PERL research groups at MSU for helpful discussions. This work is supported by Lyman Briggs College and the physics department at MSU.}
\bibliographystyle{apsrev4-1}

\bibliography{mendeley-edited}
\end{document}